\documentclass[journal]{IEEEtran}
\usepackage{graphicx}
\usepackage{subfigure}
\usepackage{caption}
\usepackage{mathptm}
\usepackage{amsmath}
\usepackage{booktabs}
\usepackage{multirow}
\usepackage{url}
\usepackage{algorithmicx}

\ifCLASSINFOpdf

\else

\fi

\begin{document}

\title{DRAMNet: Authentication based on Physical Unique Features of DRAM Using Deep Convolutional Neural Networks}

\author{Nima~Karimian,~\IEEEmembership{Member,~IEEE,}
        Fatemeh~Tehranipoor,~\IEEEmembership{Member,~IEEE,}
        Nikolaos~Anagnostopoulos,~\IEEEmembership{Student~Member,~IEEE,}
        and~Wei~Yan,~\IEEEmembership{Student~Member,~IEEE}% <-this % stops a space
\thanks{N. Karimian is with the Department
of Computer Engineering, San Jose State University, San Jose,
CA, 95192.
 Email: (nima.karimian@sjsu.edu)}% <-this % stops a space
%\thanks{Manuscript received ???, ???; revised ???, ???.}
}

%\markboth{Journal of \LaTeX\ Class Files,~Vol.~???, No.~?, ???~???}%

\maketitle

\begin{abstract}
Nowadays, there is an increasing interest in the development of Autonomous Vehicles (AV). However, there are two types of attack challenges that can affect AVs and are yet to be resolved, i.e., sensor attacks and vehicle access attacks. This paper, to the best of our knowledge, is the first work that proposes a novel authentication scheme involving DRAM power-up unique features using deep Convolutional Neural Network (CNN), which can be used to implement secure access control of autonomous vehicles. Our approach consists of two parts. First, we convert raw power-up sequence data from DRAM cells into a two-dimensional (2D) format to generate a DRAM image structure. Second, we apply deep CNN to DRAM images, in order to extract unique features from each memory to classify them for authentication. To evaluate our proposed approach, we utilize data from three Commercial-Off-The-Shelf (COTS) DRAMs taken under various environmental and other conditions (high/low temperature, high/low supply voltage and aging effects). Based on our results, our proposed authentication method ``DRAMNet'' achieves  98.63\% accuracy and 98.49\% precision. In comparison to other state-of-the-art CNN architectures, such as the AlexNet and VGGNet models, our DRAMNet approach fares equally well or better than them.
\end{abstract}


\IEEEpeerreviewmaketitle

\section{Introduction}

Autonomous vehicles are increasingly getting popular with developments in sensor technology as well as controls and communication protocols. Security is one of the key aspects to be considered in the design of AV components and systems, which have to communicate with each other and with similar systems in surrounding vehicles to ensure collision-free movement. With Internet of Things (IoT) being an integral part of AVs, it is necessary to implement robust techniques of authentication to prevent hacking of these systems, such as ~\cite{nie2017free}, allowing secure access to the black box and ensuring that counterfeit components are detected accurately during assembly. Similar concerns can be seen in other applications. Delivery drones are Unmanned Aerial Vehicles (UAVs) utilized to transport packages, food, medicine, or other goods. With high demand for a prompt and efficient delivery, a drone delivery system can be an effective solution for timely deliveries and especially for emergency management. However, current delivery drone systems lack crucial security functions. Drones may have to operate in unsupervised hostile areas, and therefore be vulnerable to physical capture in addition to conventional cyber attacks. Security solutions and methods need to be cost-efficient in terms of resource usage to be suitable for resource-limited AV and UAV systems. Among various techniques to enhance the security, Physical Unclonable Functions (PUFs) are very popular since they are easy to implement, hard to predict, and difficult to duplicate. The idea of a PUF was first proposed by Gassend et al. in~\cite{devadas}. They developed the first silicon PUFs through the use of intrinsic Process Variability (PV) in deep sub-micrometer Integrated Circuits (ICs). They used the intrinsic PV of the manufacturing of silicon devices to produce unique, random and unclonable digital responses and called it a PUF. Generally, a PUF is a function that is embodied by a physical device, which maps inputs to outputs, creating challenge-response pairs (CRPs). PUF is a completely new technology for the protection of credentials.

In this work, we focus on a new authentication method that is based on the intrinsic properties of DRAM power-up values using deep CNN, which we call DRAMNet. To do this, we have considered three COTS DRAM memories and tested their power-up values under various environmental conditions i.e. normal condition, high temperature, low temperature, high voltage, low voltage and aging. Besides the traditional PUFs that require CRPs to authenticate the chip, our proposed scheme works based on a deep learning classifier scheme using DRAM images. DRAM images are constructed using the 2D structure of the memory cells at power up. Then, we apply CNN to extract unique features from each DRAM in order to distinguish each device.  In this paper, we describe the notion of physical unique features and argue that this new method of authentication can be implemented using DRAM power-up measurements. 

\textbf{Contributions: }Specifically, this paper makes the following contributions:\\
1). We measure the power-up values of three COTS DRAMs under different environmental condition (normal condition, high temperature, low temperature, high voltage, low voltage, and aging).\\
 2). We convert the raw power-up sequence data from DRAM cells into a 2D format and then generate DRAM images.\\
 3).  We apply a deep convolutional neural network scheme to the DRAM images to classify each device for authentication purposes.\\
 4).  We examine the quality of the proposed DRAMNet model based on various metrics such as F-score, Precision, Recall and Accuracy; followed by a comparative evaluation of the proposed model against two other CNN models, namely AlexNet and VGGNet.\\
 5). We study the applicability of our DRAMNet scheme, in such applications as autonomous vehicles and unmanned aerial vehicles.

\textbf{Paper Organization: }The rest of the paper is structured as follows. We provide a literature review of previous works in Section 2. DRAM organization and properties are described in Section 3. Section 4 presents our proposed DRAMNet technique in detail. Experimental results and validation of the DRAMNet scheme is demonstrated in Section 5. Section 6 discusses DRAMNet applications in the real world. Finally, concluding remarks and future works are given in Section 7.

\section{Literature Review}
In recent years, security primitives based on the physical characteristics of DRAM have been examined in considerable detail. In particular, recent works have investigated the potential of a number of characteristics of DRAM cells to serve as the basis for the implementation of such security primitives as Physical Unclonable Functions (PUFs) and true random number generators~\cite{overview}. More specifically, the power-up values of the DRAM cells~\cite{drampuf1}, their data remanence and data retention characteristics~\cite{dramtrng1,drampuf2}, as well as their latency variations~\cite{drampuf3,drampuf4}, are all physical characteristics of DRAM modules that have been used in order to implement security primitives. Finally, we also observe that even unwanted properties of DRAM, such as its susceptibility to row hammering~\cite{cryptography2030013}, have been utilized for the implementation of DRAM-based security primitives.

However, although a number of related works regarding the usage of DRAM as a security primitive exist, to the best of our knowledge, this is the first work to connect the unique characteristics of DRAM cells with the emerging field of machine learning. We note that previous works have most often utilized machine learning as a way to attack the so-called ``strong'' PUFs~\cite{mlattack3,mlattack2,mlattack1,mlattack4} that were based on the physical variations of delay elements. Nevertheless, one of the very first works to combine the field of machine learning with the idea of security based on unique physical characteristics, actually proposed taking advantage of the information that cannot be revealed through machine learning, in order to provide secure key storage based on the physical characteristics used to implement ``strong'' PUFs~\cite{mlpuf1}. Additionally, pattern matching, which is a concept related to machine learning, has been proposed as an efficient way of implementing secure key generation using the physical characteristics of memory cells~\cite{pmpuf}. Finally, we also note that the nonlinear behavior of cellular neural networks itself has been employed in the past in order to create a PUF~\cite{cnnpuf}.

\section{DRAM Description and Properties}

\begin{figure*}
\center
\includegraphics[width=0.87\linewidth]{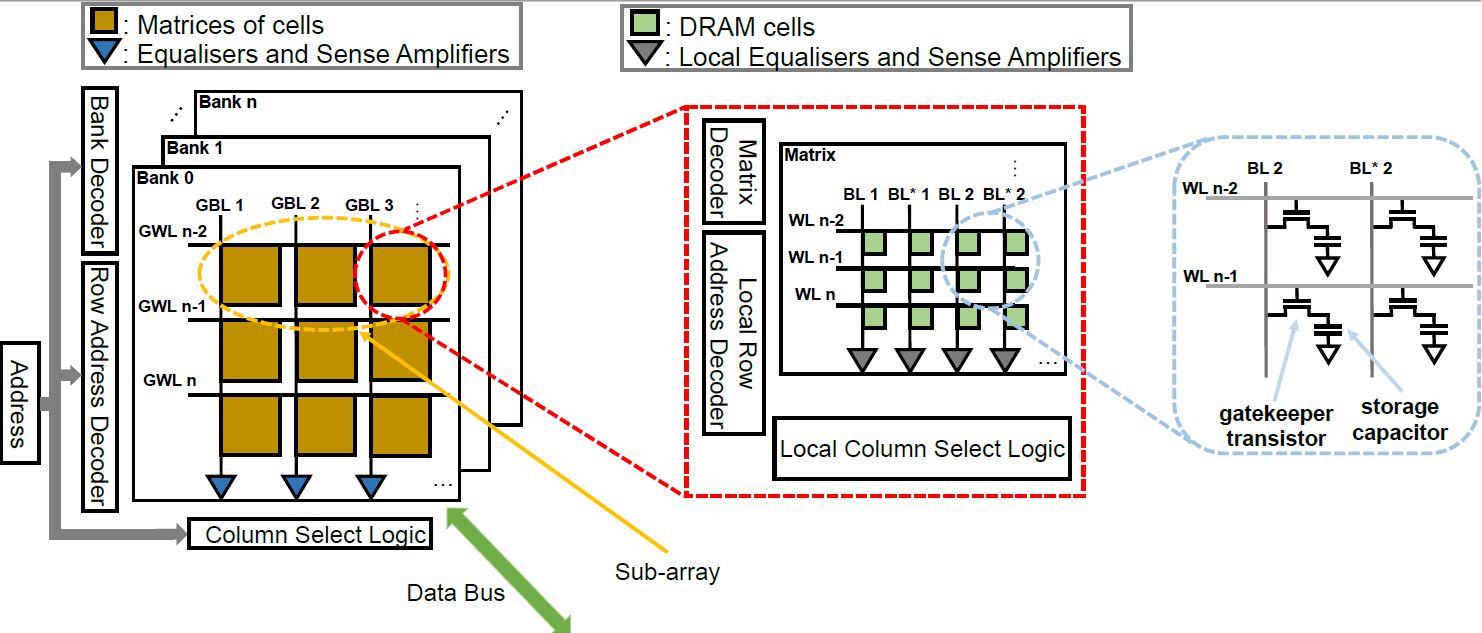}
 \caption{Internal DRAM organization. DRAM is organized in banks, each of which contains matrices of cells. Each cell consists of an access transistor and a storage capacitor.}
\label{fig:dramcells}
\end{figure*}

DRAM is a memory module that is most often an inherent component of modern computer systems. Therefore, DRAM-based security primitives tend to be highly cost-efficient and practical. Additionally, as contemporary DRAM modules are of a significantly large size (usually of some gigabytes), they allow for the extraction of a relatively large amount of unique features, based on their physical characteristics. For example, security primitives based on the power-up values of the DRAM cells offer a significantly larger amount of entropy than the security primitives that are based on the power-up values of SRAM cells, due to the significantly larger size of DRAM modules~\cite{drampuf1}. In general, DRAMs seem to be the most common type of memory in use today, can hold more data than SRAM and are significantly more cost-efficient to manufacture. In comparison, SRAMs require four times the amount of space needed by DRAMs, in order to store a given amount of data.

DRAM modules tend to be organized in a complex way, consisting of banks that incorporate arrays of several matrices of cells. For reasons of simplicity, we chose not to demonstrate this complex layout, but focus on a single matrix of cells, as shown in Fig.~\ref{fig:dramcells}. Each individual DRAM cell consists of a gatekeeper transistor and a storage capacitor, as shown in Fig.~\ref{fig:dramcells}, and stores a single bit, based on the charge state of its capacitor. In a matrix of cells, all the gates of the transistors of DRAM cells of the same row are connected to a single WordLine (WL), which enables access to all the cells of that row at the same time. In a similar fashion, the cells of each column of a matrix of cells are all connected to a single BitLine (BL or BL*), which is used to gain access to the charges of their capacitors. As each wordline allows access to all the cells of a row at once, the logical values of all the cells of a single row can be read or written at the same time using the bitlines. For this reason, adjacent bitlines are usually working as a pair, so that when one bitline (BL) is charged or discharged, the other (BL*) is, respectively, discharged or charged. Reading or writing a particular cell, or row of cells, is done by charging up the corresponding wordline and measuring the difference in the charge of the corresponding bitline(s) through sets of equalizers and sense amplifiers that connect adjacent bitlines. The logical value of a cell is determined based on whether the charge of its corresponding capacitor is above or below a certain threshold voltage value.
Additionally, as the capacitors are not ideal, the charge stored in them slowly leaks away, therefore necessitating their frequent recharging, through an operation known as {\em DRAM refresh}. Due to this operation, this type of RAM is called {\em Dynamic} in contrast to {\em Static} RAM (SRAM), which does not need to be refreshed. Furthermore, we also observe that, when different cells are charged, they may signify either a logical ``1'' (true cells -- connected to BL) or a logical ``0'' (anti-cells -- connected to BL*).

Finally, we note that the values of DRAM cells may be affected by various environmental conditions, such as temperature and voltage variations, as well as the effects of aging. Therefore, for the purposes of this work, we have stressed three commercial COTS DRAM modules under high/low temperature, high/low supply voltage and aging, by applying the techniques described in~\cite{drampuf1,dramaging1} and by using a Temptronic TP04100A ThermoStream Thermal Inducting System, in the setup shown in Fig.~\ref{fig:thermostream}. The Thermostream system delivers controlled temperature with speed and precision to devices and modules for thermal cycling and testing. It has, therefore, been used not only to test the effects of temperature variations from 0 C to 80 C on the power-up values of the DRAM cells, but also to accelerate their aging.

\section{Our Proposed DRAMNet Technique}
\subsection{DRAMNet Advantages}
The advantages of DRAMNet method are as follows:\\
1). The enrollment time required is much less than that required for the enrollment of CRPs for a PUF-based authentication method.\\ 
2). The number of samples required for enrollment are significantly fewer than those required for the enrollment of a ``strong'' PUF that provides multiple CRPs and the CNN approach does not require an error correction mechanism, unlike ``weak'' PUFs, which only provide a noisy CRP and, therefore, require error correction.\\
3). Since our DRAMNet scheme is based on deep CNN architecture, we can adapt pre-trained state-of-the-art models such as AlexNet and VGGNet to decrease the enrollment period.\\
4). DRAMNet can reduce the matching and storage complexity in the system. For example, ``strong'' PUFs need to store a large number of CRPs in a database for matching, while DRAMNet needs only one set of DRAM power-up values to be embedded into a compact template.

\begin{figure}[!t]
\includegraphics[width=\columnwidth]{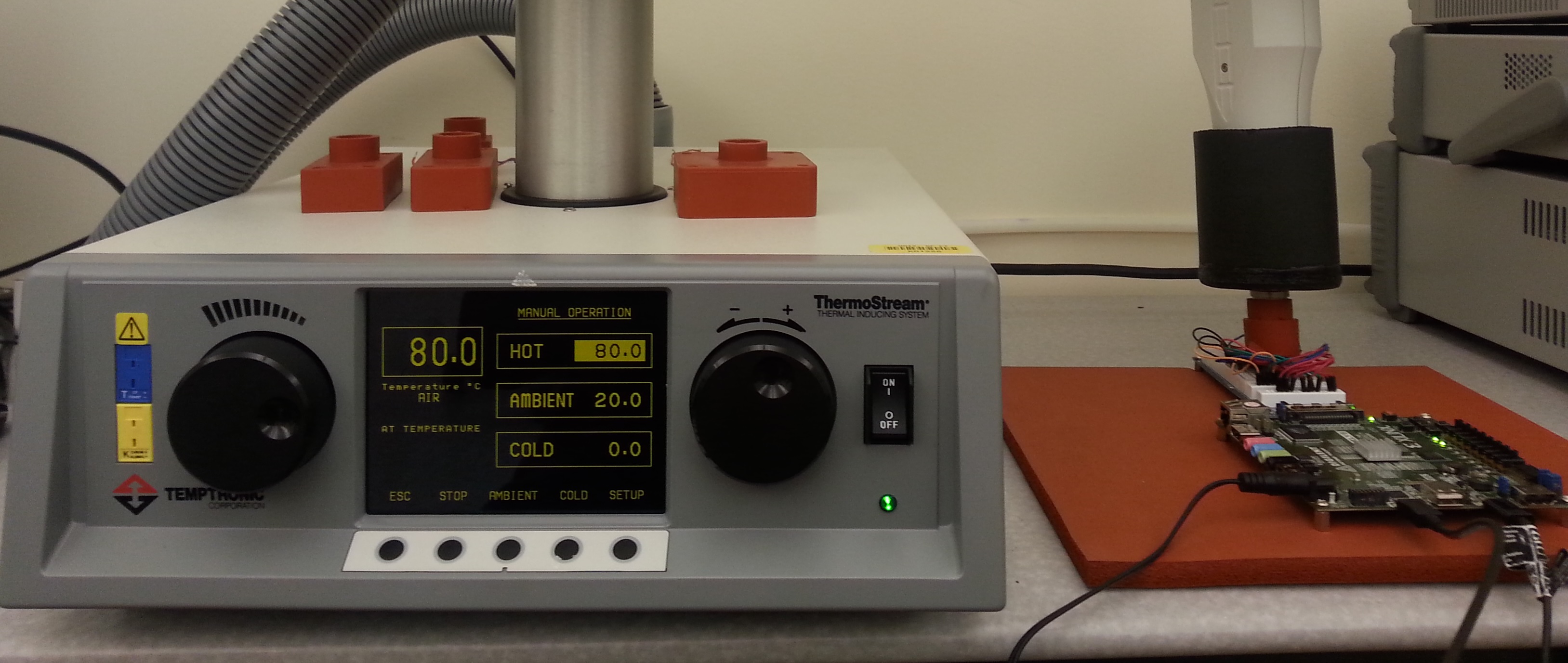}
 \caption{DRAM power-up experiments under various environmental conditions; Thermostream system (left side) and FPGA setup with DRAM (right side).}
\label{fig:thermostream}
\vspace{-10pt}
\end{figure}

\subsection{DRAM Cells 2D Structure and Images}
\begin{figure}[t]
\includegraphics[width=0.9\linewidth]{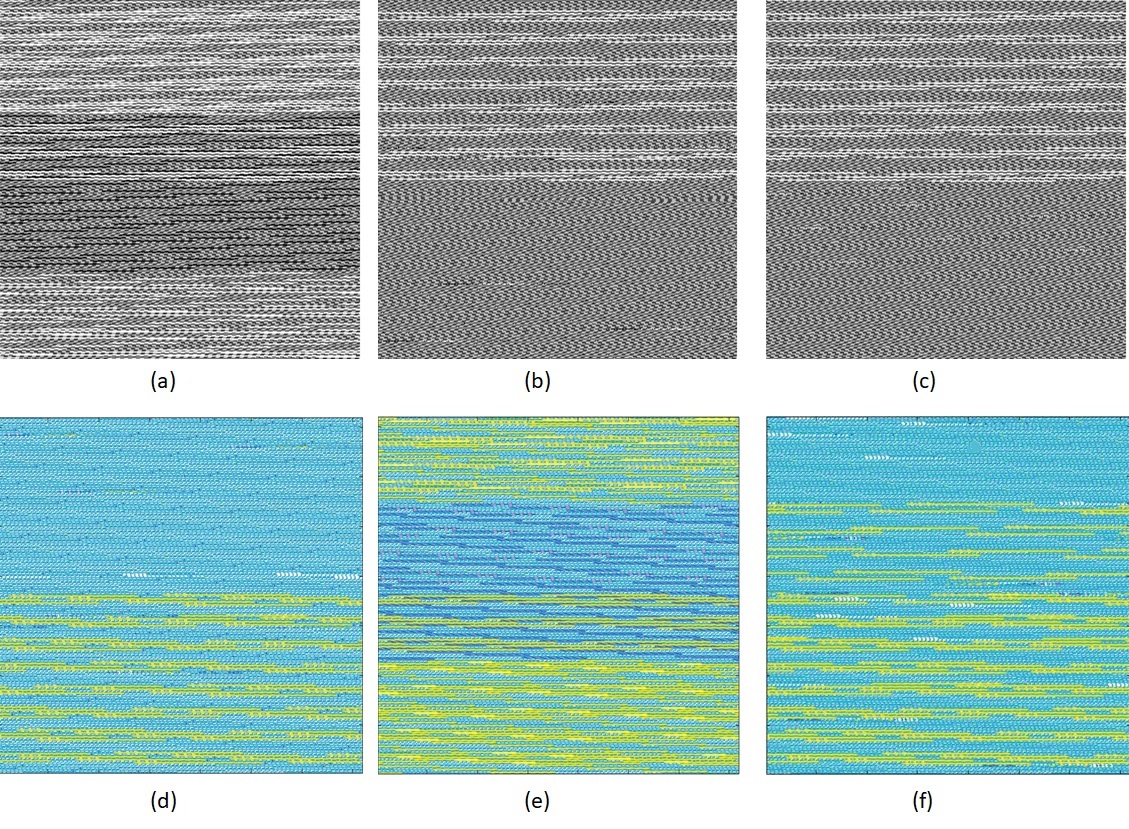}
 \caption{Converted images corresponding to each DRAM (a--c), contour plots for the DRAM images (d--f).}
\label{DRAMimage}
\vspace{-10pt}
\end{figure}

In this work, we generate unique features from DRAM modules, in order to authenticate a device. The advantage of using memory as a source of authentication is that DRAM is present on most computer systems and can be used for generating device-specific signatures without requiring any additional circuitry or hardware. In our experiments, we utilized three 1-Mbit DRAMs, which we call them DRAM1, DRAM2, and DRAM3. The power-up values of each DRAM are measured under both normal condition and various environmental and other variations, such as high voltage, low voltage, high temperature, low temperature and aging effects. Fig.~\ref{fig:thermostream} shows our experimental setup consisting of a Spartan 6 FPGA board and DRAM, on the right side of the picture, and our Temptronic TP04100A ThermoStream thermal inducting system, which was used for testing memories, on the left side of the picture. After capturing raw cell power-up values, we convert our binary data into a two-dimensional structure, consisting of multiple rows and columns. For example, for each measurement of our 1-Mbit DRAMs, a 2D structure of $1024$ rows by $1024$ columns is created. We then use these 2D structures as inputs to produce DRAM images for the deep CNN architecture. The reason of using 2D structures is their suitability for deep CNN, since one-dimensional sequence data downgrade the performance of the authentication.  We then transform these 2D structures of power-up measurement into gray-scale images using MATLAB functions, as shown in Fig.~\ref{DRAMimage}. The proposed CNN architecture is discussed in the following subsection. 

\begin{figure*}
\center
\includegraphics[width=0.87\linewidth]{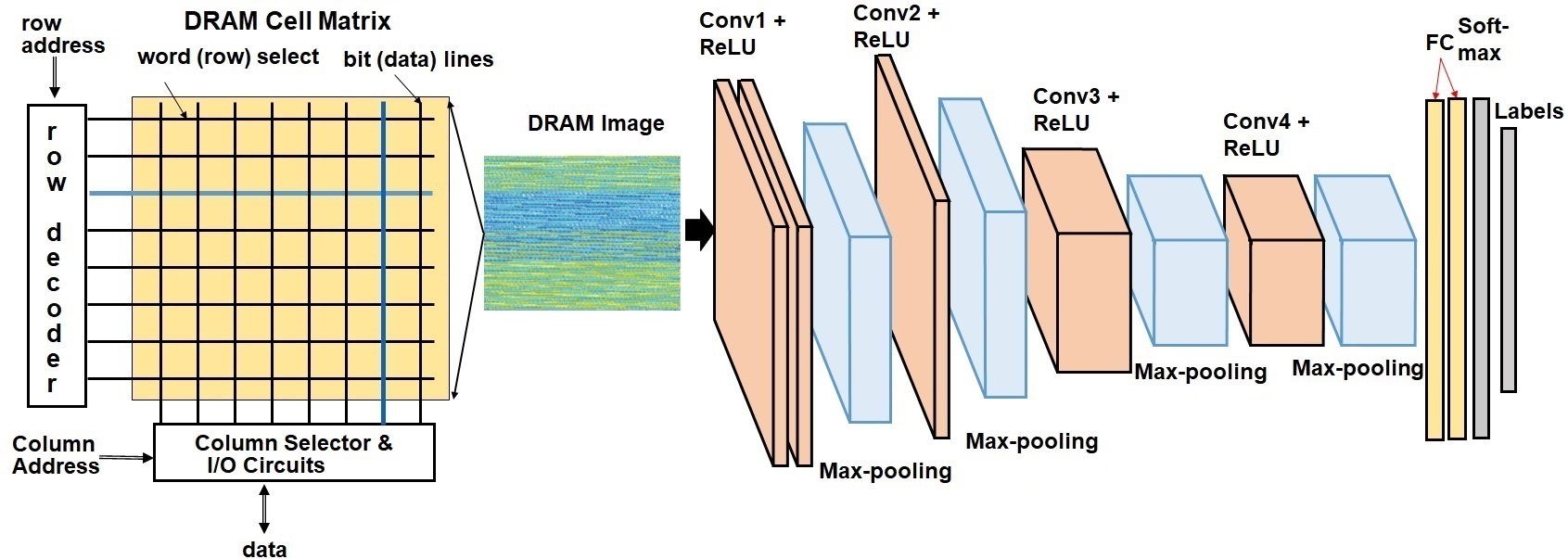}
 \caption{DRAMNet authentication network. DRAMNet consists of four convolutional and two fully-connected layers.}
\label{cnnarch}
\end{figure*}

\subsection{CNN Architecture for DRAM}	
Our classification method consists of the following steps: data acquisition, DRAM data pre-processing, and CNN classification. The DRAM power-up data discussed in this paper are obtained from COTS memories. We transform every single DRAM measurement into a $1024$ x $1024$ gray-scale image since our CNN model requires two-dimensional image as an input. Fig.~\ref{cnnarch} shows the procedures utilized in the designed CNN model. Deep learning models generally contain input, hidden, and output layers. Our designed CNN model can be divided into inputs (DRAM images), convolution methods, pooling layers for reducing the size of images, fully-connected layers for classification, and output layers (labels for each DRAM). First, each DRAM image is entered into an input layer to be then classified. Second, the convolution layer is connected to the physical unique features of DRAM image and calculates the dots product between this connected area and its own weighted value. Thirdly, the pooling layer performs downsampling per dimension and outputs the decreased volume. In the fully-connected layer, all nodes are interconnected, and the result of each node is calculated by the matrix multiplication of the weight and adding a bias to it. Finally, in the output layer, all classes are converted into a probability via the Softmax function and are classified according to the highest probability. The architecture of our CNN model is shown in Table~\ref{tab:tab1}. Additional details regarding the CNN are discussed below:

\noindent \textbf {Regularization:}
In order to reduce the overfitting in the training phase, regularization is applied, through what is typically referred to as L1 and L2 normalizations~\cite{goodfellow2016deep}. In this paper, we applied L2 regularization as indicated in Eq.~\ref{eq:Regularization}: 

\begin{equation}
\underset{w}{\mathrm{argmin}} = \sum_{j} ( t(X_{j})-\sum_{i}\omega_{i}h_{i}(X_{j}))^2 +\lambda \sum_{i=1}{k}|\omega_{i}|
\label{eq:Regularization}
\end{equation} where $\omega_{i}$ is a weight vector, which parametrizes the space of functions $t(X_{j})$ that we are interested in considering. $\lambda $ indicates how much we are penalizing the norm of $\omega$. Additionally, we used dropout and batch normalization methods~\cite{srivastava14a} for deep CNN models to avoid overfitting. In this paper, we applied batch normalization before the activation function and after the convolution layer. Dropout with a probability of $0.5$ was applied after the batch normalization layer of the fully-connected block.

\noindent \textbf {Cost function:}
To evaluate the effective performance of training neural networks, a cross-entropy function is used to minimize the difference between the given training DRAM image and the desired output. The relevant cost function is defined as Eq.~\ref{eq:cost}:

\begin{equation}
C= -\frac{1}{N} \sum [d \ln{y}+(1-d)\ln(1-y)] 
\label{eq:cost}
\end{equation} where $n$ is the batch size, d is the desired output, and $y$ is the actual value from the output layer. We apply a gradient decent optimizer function with an initial learning rate of 0.01 and 0.9 decay per 500 steps to minimize the cost function. Before each epoch, the training set is shuffled and each mini-batch (16 images) is then picked, thus ensuring each image is used only once in an epoch.

\noindent \textbf {Optimized CNN classifier architecture:}
Considering the above procedure, we designed our CNN model for DRAM-based authentication using CNN classification. The main structure of our CNN model is similar to VGGNet, which optimizes various functions to reduce overfitting and improve classification accuracy. Table~\ref{tab:tab1} and Fig.~\ref{cnnarch} describe the detailed architecture of our CNN model. The proposed CNN model is compared to AlexNet~\cite{krizhevsky2012imagenet} and VGGNet~\cite{simonyan2014very} in the following section.

\begin{table}[]
%\small
\caption {Architecture of the Proposed CNN Model} \label{tab:tab1} 
\begin{tabular}{llcccl}
\hline
 & Type & \multicolumn{1}{l}{Kernel size} & \multicolumn{1}{l}{Stride} & \multicolumn{1}{l}{\#Kernel} & Input size \\ \hline
Layer1 & Conv2D & 3 x 3 & 1 & 3 & 1024 x 1024 x 1 \\
Layer2 & Conv2D & 3 x 3 & 1 & 64 & 1024 x 1024 x 3 \\
Layer3 & Pool & 2 x 2 & 2 &  & 1024 x 1024 x 64 \\
Layer4 & Conv2D & 3 x 3 & 1 & 128 & 512 x 512 x 64 \\
Layer5 & Pool & 2 x 2 & 2 &  & 180 x 512 x 512 \\
Layer6 & Conv2D & 3 x 3 & 1 & 192 & 256 x 256 x 128 \\
Layer7 & Pool & 2 x 2 & 2 &  & 128 x 128 x 192 \\
Layer8 & Full &  &  & 2048 & 128 x 128 x 192 \\
Layer9 & Full &  &  & 2048 & 2048 \\
Layer10 & Out &  &  & 3 & 2048 \\ \hline
\end{tabular}
\end{table}

\section{Experimental Setup and Validation}
\label{sec:experiments}

\noindent \textbf {Experimental Setup:}
The proposed classifier and two other CNN models (AlexNet and VGGNet) were implemented in the Keras Python library with the TensorFlow backend; an open source software library for deep learning launched by Google. For our experiments, we have tested 180 measurements in total from three DRAMs (DRAM1, DRAM2, DRAM3) under different operating conditions. We specified 60\% of the total measurement data (DRAM images) for training DRAMNet and the rest for the testing/evaluation phase. Note that for training data, we ensured that the percentage of DRAM images selected from each particular DRAM was equal.

\noindent \textbf {Training of the CNN:}
To train the CNN, Stochastic Gradient Descent (SGD) with momentum~\cite{kushner2003stochastic} is applied for optimization. SGD reduces the time needed for training, by using only a batch of the training instances in each iteration. Therefore the computed error value is biased with respect to the optimum, but can be performed much faster. Each iteration over the entire training set (epoch) requires multiple iterations over the small batches. In addition, the Adam optimizer function~\cite{kingma2014adam} with $0.001$ starting learning rate and 0.9 decay rate is used. Since our network is used as a classifier for DRAM authentication with three classes, the output layer contains one neuron per class. Finally, we have applied the SoftMax function to convert these classes into probabilities, where SoftMax is the probability of the input to belong to each DRAM.

\textbf {Evaluation Metrics for Comparison:}
To evaluate the effective performance of our proposed system, the following four metrics are utilized: accuracy, precision, recall, and F-score. Eq.~\ref{eq:metric} shows how accuracy, precision, recall, and F$_1$-score are derived. These four metrics are the most frequently used performance indicators for machine learning models.

\begin{figure*}[t]
\begin{tabular}{cccc}
\hspace*{-1.5em} \includegraphics[width=0.27\linewidth]{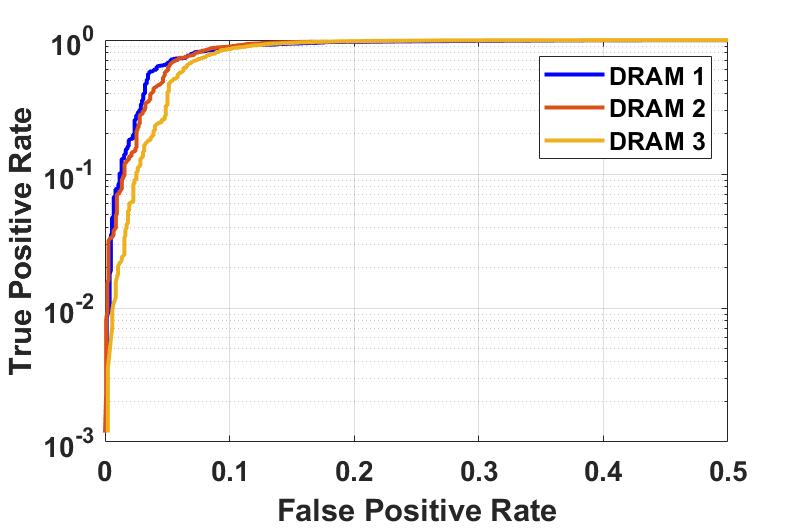}\hspace*{-2.2em} &
\includegraphics[width=0.27\linewidth]{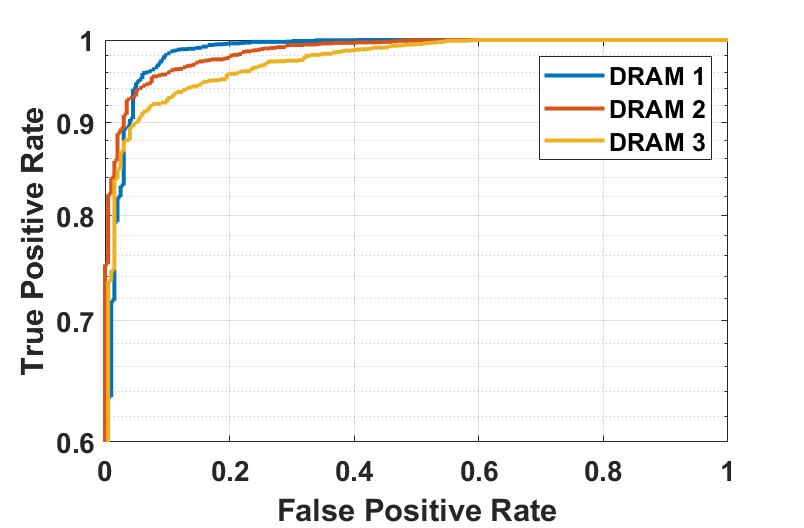}\hspace*{-2.2em} &
\includegraphics[width=0.27\linewidth]{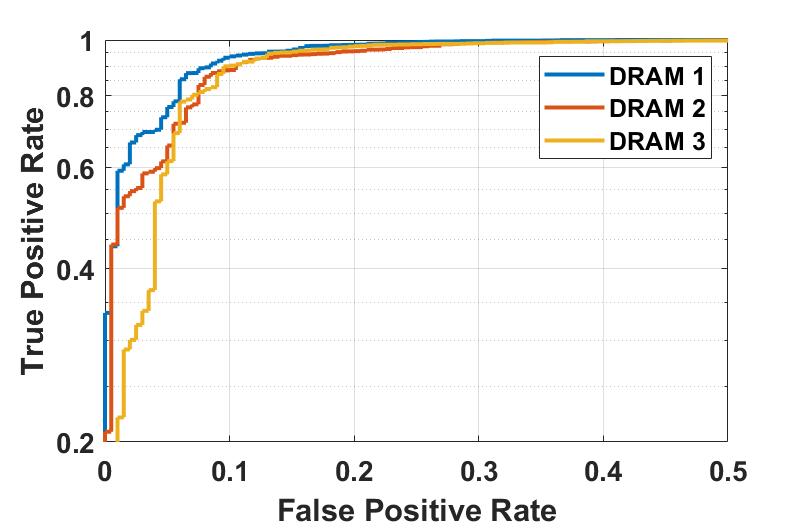}\hspace*{-2.2em} &
\includegraphics[width=0.27\linewidth]{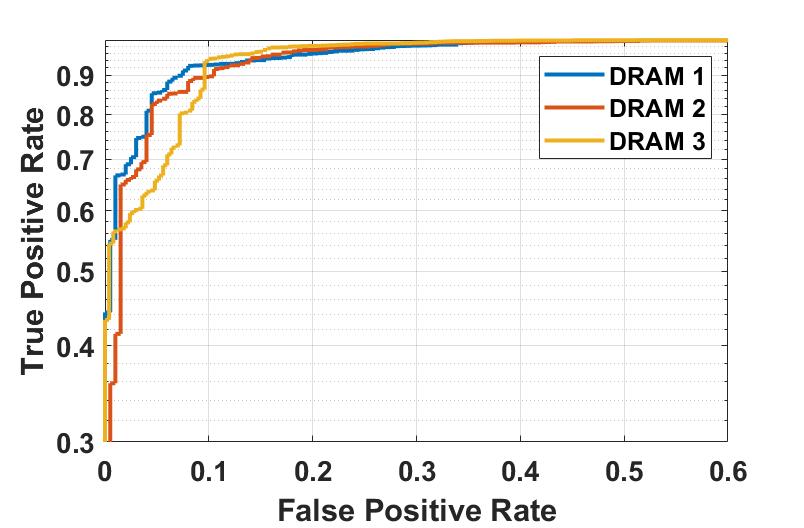}\hspace*{-2.2em}\\
(a) & (b) & (c) & (d)
\end{tabular}\vspace{-5pt}
\caption{ROC curves of different DRAMs for (a) DRAMNet based on the SGD optimizer, (b) DRAMNet based on the Adam optimizer, (c) VGGNet based on the Adam optimizer, and (d) AlexNet based on the Adam optimizer.}
\label{roc}
\vspace{-5pt}
\end{figure*}

\begin{table*}[t]
\caption {Summarized Performance Results} \label{tab:results} 
\center
\begin{tabular}{llcccccccc}
\hline
Method & Type & \multicolumn{2}{c}{Accuracy \%} & \multicolumn{2}{c}{Recall \%} & \multicolumn{2}{c}{Precision \%} & \multicolumn{2}{c}{F-score \%} \\ \hline
 &  & SGD & \multicolumn{1}{l}{Adam} & SGD & \multicolumn{1}{l}{Adam} & SGD & \multicolumn{1}{l}{Adam} & \multicolumn{1}{l}{SGD} & \multicolumn{1}{l}{Adam} \\ \cline{3-10} 
\multirow{2}{*}{DRAMNet} & Original & 91.23 & 94.35 & 89.56 & 94.21 & 89.29 & 94.17 & 88.42 & 94.19 \\
 & Augmented & 95.52 & 98.63 & 95.44 & 98.29 & 95.50 & 98.49 & 95.47 & 98.40 \\
\multirow{2}{*}{AlexNet} & Original & 91.89 & 96.49 & 91.52 & 95.39 & 89.37 & 96.41 & 90.43 & 95.89 \\
 & Augmented & 96.93 & 98.69 & 96.81 & 98.23 & 96.66 & 98.51 & 96.73 & 98.36 \\
\multirow{2}{*}{VGGNet} & Original & 91.49 & 95.44 & 91.07 & 94.89 & 88.73 & 94.73 & 89.88 & 94.97 \\
 & Augmented & 94.72 & 97.20 & 94.68 & 96.95 & 94.34 & 96.88 & 94.50 & 96.91 \\ \hline
\end{tabular}
\end{table*}

\begin{equation}\label{eq:metric}
\begin{split}
Accuracy &= \frac{TP + TN}{TP + TN + FP + FN} \\
Precision &= \frac{TP}{TP + FP} \\
Recall &= \frac{TP}{TP + FN} \\
F_1-Score &= 2\times \frac{Precision\times Recall}{Precision+Recall}
\end{split}
\end{equation} where TP, TN, FP, and FN represent the number of true positives, true negatives, false positives, and false negatives,
respectively. Moreover, we have also calculated the Receiver Operation Characteristic (ROC) curves. This curve shows the trade-off between False Positive Rate (FPR) and True Positive Rate (TPR). FPR refers to the rate at which a classifier incorrectly matches the non-matching data (outlier) to the target class. TPR refers to the rate that a classifier correctly matches the matching data (target) to the target class.

\textbf{Classification Results and Discussion:}
Table~\ref{tab:results} presents the summarized performance results of three CNN models, where DRAMNet refers to our CNN model and AlexNet and VGGNet refer to other existing CNN architectures. First, the training data was conducted without augmentation. In this case, the DRAMNet CNN model achieved 91.23\% average accuracy, 89.56\% recall, 89.29\% average precision and 88.42\% F-score based on the SGD optimizer. Since our DRAMNet pool is not large enough to achieve the optimal performance, training DRAMNet data can be enlarged by augmenting the DRAM images which results in higher authentication accuracy. Data augmentation is one of the advantages of applying images as input data into the deep learning technique. We augmented DRAM images by cropping each image in six different ways at the training model. By using this augmentation, our DRAMNet model achieved more than 6\% higher accuracy, recall, precision, and F-score. Surprisingly, the authentication performance is increased by applying the Adam optimizer technique for all the models. Based on the results presented in Table~\ref{tab:results}, the AlexNet CNN model exhibits the best accuracy, recall, and sensitivity precision when we train the data without augmentation based on the SGD optimizer. However, the accuracy for AlexNet, VGGNet, and DRAMNet improved by a factor of $7.4$, $6.2$, and $8.1$, respectively when data augmentation was applied. The discrimination of recall between the DRAMNet method and first highest of the other ones is only 0.06\% while the discrimination of precision of the proposed method and the highest of the other ones is 0.1\%. In general, we note that data augmentation and the use of the Adam optimizer technique helps all CNN models achieve their highest accuracy, recall, precision, and F-score. In particular, the average accuracy of DRAMNet for both original and augmented data is 94.35\% versus 98.63\% using the Adam optimizer. In addition, the average accuracy of AlexNet and VGGNet is 98.69\% and 97.20\%, respectively, for augmented data and using the Adam optimizer. Therefore, we observe that the DRAMNet and AlexNet approaches obtain the best classification performance overall. In order to analyze further the performance of the three CNN models, we also study the accuracy of their authentication, using ROC curves. The tradeoff between the TPR and FPR for each CNN model and each of the three DRAMs, are shown in the ROC curves presented Fig.~\ref{roc}, and denote the accuracy of each model. In conclusion, in comparison to the other two state-of-the-art CNN architectures, i.e. the AlexNet and VGGNet models, DRAMNet provides equally good, or even better, results.

\section{DRAMNet Applications}
DRAMNet provides an alternative approach to the traditional authentication methods. DRAM, as a low-cost and high-capacity memory, is an ideal choice to meet the constraints of embedded hardware. Additionally, as most graphics hardware can inherently support the application of CNN, SDRAM used in graphics is particularly suited for the implementation of the DRAMNet method, which utilizes CNN to distinguish the unique DRAM power-up values of different DRAM modules. Therefore, DRAMNet can be used to identify and authenticate different devices, without requiring additional resources. In particular, DRAMNet can serve to implement secure access control for AVs and UAVs. In this case, access is only provided to entities that provide DRAM-based images that DRAMNet validates as matching its training data. Another application involves UAVs that deliver goods. In this case, UAVs will only communicate with entities authenticated though DRAMNet. A client making a purchase must send a number of DRAM-based images to the host service (shop), which are then used to train the corresponding UAV's DRAMNet, so that it can identify, authenticate and communicate with that particular client, by using its DRAMNet implementation to match future DRAM-based images sent by this client. However, DRAMNet applications may be vulnerable to data retrieving and man-in-the-middle or replay attacks. In order to overcome such potential issues, we suggest that images received should no longer be stored after being used for enrollment or identification/authentication and that all communication between entities use asymmetric encryption.

\section{Conclusions and Future Work}
In this paper, we have presented a novel authentication method based on DRAM unique features using deep CNN, ``DRAMNet''. Our results indicate that the proposed method has the potential to be used to secure access control in such applications as autonomous vehicles and drones. To evaluate our approach, we utilized data from three COTS DRAM modules taken under various environmental and other conditions, such as high/low temperature, high/low supply voltage and aging effects. Based on our results, DRAMNet achieved  98.63\% accuracy and 98.49\% precision. In comparison to other state-of-the-art CNN architectures, such as the AlexNet and VGGNet models, DRAMNet provides equally good, or even better, results. In the future, we will take generate a larger number of DRAM measurements, in order to increase the training pool size of the CNN and improve the accuracy of authentication. We also intend to try these experiments using other types of DRAM modules and characteristics.

\ifCLASSOPTIONcaptionsoff
  \newpage
\fi

\bibliographystyle{IEEEtran}

%\bibliography{IEEEexample}

\begin{thebibliography}{}
\providecommand{\url}[1]{#1}
\csname url@samestyle\endcsname
\providecommand{\newblock}{\relax}
\providecommand{\bibinfo}[2]{#2}
\providecommand{\BIBentrySTDinterwordspacing}{\spaceskip=0pt\relax}
\providecommand{\BIBentryALTinterwordstretchfactor}{4}
\providecommand{\BIBentryALTinterwordspacing}{\spaceskip=\fontdimen2\font plus
\BIBentryALTinterwordstretchfactor\fontdimen3\font minus
  \fontdimen4\font\relax}
\providecommand{\BIBforeignlanguage}[2]{{%
\expandafter\ifx\csname l@#1\endcsname\relax
\typeout{** WARNING: IEEEtran.bst: No hyphenation pattern has been}%
\typeout{** loaded for the language `#1'. Using the pattern for}%
\typeout{** the default language instead.}%
\else
\language=\csname l@#1\endcsname
\fi
#2}}
\providecommand{\BIBdecl}{\relax}
\BIBdecl

\end{thebibliography}


\begin{thebibliography}{10}

\bibitem{nie2017free}
S.~Nie, L.~Liu, and Y.~Du, ``Free-fall: Hacking tesla from wireless to can
  bus.'' {Black Hat USA, Mandalay Bay, Las Vegas, NV, USA}, 2017.

\bibitem{devadas}
B.~Gassend, D.~Clarke, M.~van Dijk, and S.~Devadas, ``Silicon physical random
  functions,'' in {\em Proceedings of the 9th ACM Conference on Computer and
  Communications Security}, CCS '02, (New York, NY, USA), pp.~148--160, ACM,
  2002.

\bibitem{overview}
N.~A. Anagnostopoulos, S.~Katzenbeisser, J.~Chandy, and F.~Tehranipoor, ``An
  overview of dram-based security primitives,'' {\em Cryptography}, vol.~2,
  no.~2, 2018.

\bibitem{drampuf1}
F.~Tehranipoor, N.~Karimian, W.~Yan, and J.~A. Chandy, ``Dram-based intrinsic
  physically unclonable functions for system-level security and
  authentication,'' {\em IEEE Transactions on Very Large Scale Integration
  (VLSI) Systems}, vol.~25, pp.~1085--1097, March 2017.

\bibitem{dramtrng1}
F.~Tehranipoor, W.~Yan, and J.~A. Chandy, ``Robust hardware true random number
  generators using dram remanence effects,'' in {\em 2016 IEEE International
  Symposium on Hardware Oriented Security and Trust (HOST)}, pp.~79--84, May
  2016.

\bibitem{drampuf2}
A.~Schaller, W.~Xiong, N.~A. Anagnostopoulos, M.~U. Saleem, S.~Gabmeyer,
  B.~\v{S}kori\'{c}, S.~Katzenbeisser, and J.~Szefer, ``Decay-based dram pufs
  in commodity devices,'' {\em IEEE Transactions on Dependable and Secure
  Computing}, 2018.

\bibitem{drampuf3}
B.~M. S.~B. Talukder, B.~Ray, M.~Tehranipoor, D.~Forte, and M.~T. Rahman,
  ``{LDPUF:} exploiting {DRAM} latency variations to generate robust device
  signatures,'' {\em arXiv preprint arXiv:1808.02584}, 2018.

\bibitem{drampuf4}
J.~S. Kim, M.~Patel, H.~Hassan, and O.~Mutlu, ``The dram latency puf: Quickly
  evaluating physical unclonable functions by exploiting the
  latency-reliability tradeoff in modern commodity dram devices,'' in {\em 2018
  IEEE International Symposium on High Performance Computer Architecture
  (HPCA)}, pp.~194--207, Feb 2018.

\bibitem{cryptography2030013}
N.~A. Anagnostopoulos, T.~Arul, Y.~Fan, C.~Hatzfeld, A.~Schaller, W.~Xiong,
  M.~Jain, M.~U. Saleem, J.~Lotichius, S.~Gabmeyer, J.~Szefer, and
  S.~Katzenbeisser, ``Intrinsic run-time row hammer pufs: Leveraging the row
  hammer effect for run-time cryptography and improved security,'' {\em
  Cryptography}, vol.~2, no.~3, 2018.

\bibitem{mlattack3}
U.~R\"{u}hrmair, F.~Sehnke, J.~S\"{o}lter, G.~Dror, S.~Devadas, and
  J.~Schmidhuber, ``Modeling attacks on physical unclonable functions,'' in
  {\em Proceedings of the 17th ACM Conference on Computer and Communications
  Security}, CCS '10, (New York, NY, USA), pp.~237--249, ACM, 2010.

\bibitem{mlattack2}
U.~Rührmair, J.~Sölter, F.~Sehnke, X.~Xu, A.~Mahmoud, V.~Stoyanova, G.~Dror,
  J.~Schmidhuber, W.~Burleson, and S.~Devadas, ``Puf modeling attacks on
  simulated and silicon data,'' {\em IEEE Transactions on Information Forensics
  and Security}, vol.~8, pp.~1876--1891, Nov 2013.

\bibitem{mlattack1}
F.~Ganji, S.~Tajik, F.~F{\"a}{\ss}ler, and J.-P. Seifert, ``Strong machine
  learning attack against pufs with no mathematical model,'' in {\em
  Cryptographic Hardware and Embedded Systems -- CHES 2016} (B.~Gierlichs and
  A.~Y. Poschmann, eds.), (Berlin, Heidelberg), pp.~391--411, Springer Berlin
  Heidelberg, 2016.

\bibitem{mlattack4}
C.~Herder, M.~Yu, F.~Koushanfar, and S.~Devadas, ``Physical unclonable
  functions and applications: A tutorial,'' {\em Proceedings of the IEEE},
  vol.~102, pp.~1126--1141, Aug 2014.

\bibitem{mlpuf1}
M.-D.~M. Yu, D.~M'Raihi, R.~Sowell, and S.~Devadas, ``Lightweight and secure
  puf key storage using limits of machine learning,'' in {\em Cryptographic
  Hardware and Embedded Systems -- CHES 2011} (B.~Preneel and T.~Takagi, eds.),
  (Berlin, Heidelberg), pp.~358--373, Springer Berlin Heidelberg, 2011.

\bibitem{pmpuf}
Z.~Paral and S.~Devadas, ``Reliable and efficient puf-based key generation
  using pattern matching,'' in {\em 2011 IEEE International Symposium on
  Hardware-Oriented Security and Trust}, pp.~128--133, June 2011.

\bibitem{cnnpuf}
T.~Addabbo, A.~Fort, M.~D. Marco, L.~Pancioni, and V.~Vignoli, ``Physically
  unclonable functions derived from cellular neural networks,'' {\em IEEE
  Transactions on Circuits and Systems I: Regular Papers}, vol.~60,
  pp.~3205--3214, Dec 2013.

\bibitem{goodfellow2016deep}
I.~Goodfellow, Y.~Bengio, and A.~Courville, {\em Deep Learning}.
\newblock MIT Press, 2016.

\bibitem{dramaging1}
F.~Tehranipoor, N.~Karimian, W.~Yan, and J.~A. Chandy, ``Investigation of dram
  pufs reliability under device accelerated aging effects,'' in {\em 2017 IEEE
  International Symposium on Circuits and Systems (ISCAS)}, pp.~1--4, May 2017.

\bibitem{srivastava14a}
N.~Srivastava, G.~Hinton, A.~Krizhevsky, I.~Sutskever, and R.~Salakhutdinov,
  ``Dropout: A simple way to prevent neural networks from overfitting,'' {\em
  Journal of Machine Learning Research}, vol.~15, pp.~1929--1958, 2014.

\bibitem{kushner2003stochastic}
H.~J. Kushner and G.~G. Yin, {\em Stochastic Approximation and Recursive
  Algorithms and Applications}, vol.~35 of {\em Stochastic Modelling and
  Applied Probability}.
\newblock Springer Science \& Business Media, 2003.

\bibitem{kingma2014adam}
D.~P. Kingma and J.~Ba, ``Adam: A method for stochastic optimization,'' {\em
  arXiv preprint arXiv:1412.6980}, 2014.

\bibitem{krizhevsky2012imagenet}
A.~Krizhevsky, I.~Sutskever, and G.~E. Hinton, ``Imagenet classification with
  deep convolutional neural networks,'' in {\em Proceedings of the 25th
  International Conference on Neural Information Processing Systems - Volume
  1}, NIPS'12, (USA), pp.~1097--1105, Curran Associates Inc., 2012.

\bibitem{simonyan2014very}
K.~Simonyan and A.~Zisserman, ``Very deep convolutional networks for
  large-scale image recognition,'' {\em arXiv preprint arXiv:1409.1556}, 2014.

\end{thebibliography}

\end{document}